\documentclass{aa}
\usepackage[dvips]{graphicx}
\usepackage{psfig}

\begin{document}
 \title{Spectroscopic distances of nearby ultracool dwarfs}

   \author{N.~Phan-Bao
       \inst{1}
       \and
       M.S.~Bessell
       \inst{2}
       }

\offprints{N.~Phan-Bao, \email{pbngoc@asiaa.sinica.edu.tw}}

\institute{Institute of Astronomy and Astrophysics, Academia Sinica.
          P.O. Box 23-141, Taipei 106, Taiwan, R.O.C.
          \and
	  Research School of Astronomy and Astrophysics, Australian National University, 
           Cotter Rd, Weston, ACT 2611, Australia
    	  }

      \date{Received / Accepted}

\abstract{We present updated results of spectroscopic follow-up observations of a sample of 45 M dwarf candidates 
identified by Phan-Bao et al. (\cite{phan-bao03}) based on the DENIS photometry and proper motion measurements.
Forty one of these are nearby late-M dwarfs (d$<$30~pc) with spectral types 
ranging from M5.0 to M8.5 computed from the spectral indices. One contaminant is probably an F-G main
sequence star reddened by intervening dust and three stars that were not observed have previous classifications 
as M dwarfs in the literature.
In this paper, we identify three M7.5, five M8.0, one M8.5 dwarf and confirm two new M8.0 dwarf members of 
the 25 pc volume. 
 }           
\titlerunning{nearby ultracool dwarfs}
\authorrunning{Phan-Bao \& Bessell}
  \maketitle


\section{Introduction}

We present updated results of spectroscopic observations of nearby red dwarf
candidates (d $<$ 30~pc; $2\leq I-J\leq3$)
from Phan-Bao et al. (\cite{phan-bao01} and 
\cite{phan-bao03}, hereafter Papers~I and II respectively). 
These late-M dwarf candidates were either found in the DENIS survey 
(Epchtein et al. \cite{epchtein97}) 
over 5700 square degrees, or cross-identified over a wider area between 
the DENIS database and the LHS or NLTT catalogs (Luyten \cite{luytena}, 
Luyten \cite{luytenb}). All were further selected by the {"Maximum Reduced 
Proper Motion"} method (hereafter MRPM, see Paper~II). 
The basic idea of this method is to use the reduced-proper-motion 
(H = M + 5 log(V$_{\rm t}$/4.74)) versus color diagram to distinguish 
between nearby ultra-cool dwarfs and more distant giants. This is done as follows:
(1) Calculate the maximum reduced-proper-motion of a red giant at a given color - this 
corresponds to the largest possible tangential velocity, V$_{\rm t}$ of about 800~kms$^{-1}$.
(2) Any object with a reduced proper motion, calculated from the DENIS photometry and our proper motion 
measurements, larger than the maximum value of a red giant at the same color must be a dwarf.

That robust selection method retrieves solar neighbourhood late-M dwarfs down to very low proper motions
(e.g., DENIS-P J1538317$-$103850, an M5.0 with $\mu$ = 20~mas/yr),
well below the 0.18~arc-sec/yr threshold of the NLTT catalog (Luyten, 1979) 
and can therefore retrieve a good fraction of the estimated 26\% nearby late-M dwarfs (d $<$ 30~pc)
missing from the NLTT catalog (Crifo et al. \cite{crifo}).

The first spectroscopic observations of 32 M dwarfs have been presented by Crifo et al. (\cite{crifo})
based on observations made at the ESO-1.52m and NTT-3.6m telescopes at La Silla, Chile in 2002 and 2003. 
In this paper, we present spectroscopic observations of the remaining candidates 
made at the MSSSO-2.3m telescope.

Section~2 describes our sample, spectroscopic observations and reductions. 
Section~3 presents the measurements of various spectroscopic indices,
spectral type and distance estimates. Section~4
discusses the chromospheric activity in M dwarfs. 
We summarize our results in the final section.

\section{Data sample and spectroscopic observations}
\subsection{Data sample}
\label{Data_sample}
\begin{table*}
   \caption{Spectral indices for the 41  nearby red dwarfs and the VB 10 standard observed at MSSSO-2.3m}
    \label{table_indices}
  $$
   \begin{tabular}{lllllllllllll}
   \hline 
   \hline
   \noalign{\smallskip}
Stars              &  NLTT/LHS   & H$_{\alpha}$ &  PC3   & CrH1  &  CrH2 &  FeH1  &  FeH2  &  CaH1 & CaH2  & CaH3 & TiO5 &  VOa  \\
                   &   name      &              &        &       &       &        &        &       &       &      &      &       \\
  (1)&(2) &(3) &(4)  &(5) &(6) &(7) &(8)  &(9)  & (10)  & (11)  &  (12)  &  (13)   \\
      \noalign{\smallskip}
\hline
J0002061$+$011536  &  LP 584-4    &  1.35   &  1.60   &  0.97   &  1.15   &  0.95   &  1.32   &  0.842   &  0.276   &  0.608   &  0.265   &  2.20   \\
J0020231$-$234605  &  LP 825-35   &  1.26   &  1.51   &  0.98   &  1.12   &  0.96   &  1.29   &  0.933   &  0.283   &  0.615   &  0.297   &  2.12   \\
J0041353$-$562112  &  ...         &  2.83   &  1.69   &  0.95   &  1.13   &  0.89   &  1.31   &  0.910   &  0.269   &  0.613   &  0.269   &  2.27   \\
J0103119$-$535143  &  ...         &  1.49   &  1.44   &  0.97   &  1.07   &  0.94   &  1.21   &  0.882   &  0.300   &  0.620   &  0.297   &  2.12   \\
J0120491$-$074103  &  ...         &  2.70   &  1.80   &  0.99   &  1.14   &  0.94   &  1.22   &  1.134   &  0.289   &  0.583   &  0.307   &  2.31   \\
J0144318$-$460432  &  ...         &  2.24   &  1.38   &  0.94   &  1.09   &  0.91   &  1.25   &  0.832   &  0.281   &  0.584   &  0.265   &  2.08   \\
J0145434$-$372959  &              &  1.03   &  1.75   &  0.94   &  1.14   &  0.88   &  1.22   &  0.801   &  0.254   &  0.540   &  0.250   &  2.23   \\
J0218579$-$061749  &  LP 649-93   &  1.23   &  1.64   &  0.98   &  1.16   &  0.91   &  1.22   &  0.798   &  0.261   &  0.612   &  0.235   &  2.22   \\
J0235495$-$071121  &  ...         &  1.25   &  1.47   &  0.99   &  1.15   &  0.96   &  1.24   &  0.840   &  0.295   &  0.584   &  0.287   &  2.11   \\
J0253444$-$795913  &  ...         &  1.60   &  1.38   &  0.97   &  1.12   &  0.98   &  1.23   &  0.795   &  0.262   &  0.566   &  0.270   &  2.05   \\
J0312251$+$002158  &  ...         &  1.11   &  1.37   &  0.99   &  1.13   &  1.00   &  1.25   &  0.643   &  0.300   &  0.582   &  0.295   &  2.10   \\
J0320588$-$552015  &  ...         &  1.27   &  1.42   &  0.97   &  1.08   &  0.96   &  1.19   &  0.848   &  0.303   &  0.610   &  0.293   &  2.09   \\
J0324268$-$772705  &              &  1.16   &  1.56   &  0.98   &  1.18   &  1.00   &  1.32   &  0.821   &  0.268   &  0.540   &  0.259   &  2.08   \\
J0413398$-$270428  &  LP 890-2    &  1.23   &  1.43   &  0.96   &  1.15   &  0.96   &  1.20   &  0.862   &  0.279   &  0.590   &  0.252   &  2.11   \\
J0517377$-$334903  &  ...         &  1.34   &  2.08   &  1.03   &  1.26   &  1.04   &  1.41   &  0.913   &  0.350   &  0.605   &  0.394   &  2.22   \\
J0520293$-$231848  &  LP 836-41   &  1.02   &  1.31   &  0.95   &  1.06   &  0.92   &  1.15   &  0.875   &  0.348   &  0.687   &  0.317   &  2.06   \\
J0602542$-$091503  &  LHS 1810    &  1.18   &  1.20   &  0.97   &  1.10   &  0.97   &  1.16   &  0.859   &  0.273   &  0.604   &  0.278   &  2.07   \\
J1021323$-$204407  &  ...         &  0.67:  &  2.06   &  1.10   &  1.22   &  1.09   &  1.32   &  1.612:  &  0.210:  &  0.528   &  0.264:  &  2.23   \\
J1021513$-$032309  &  LP 610-5    &  0.86:  &  1.42   &  1.08   &  1.11   &  1.03   &  1.25   &  0.421:  &  0.212   &  0.435   &  0.081:  &  1.96:  \\
J1106569$-$124402  &  LP 731-47   &  1.25   &  1.45   &  0.98   &  1.14   &  0.97   &  1.24   &  0.884   &  0.265   &  0.565   &  0.258   &  2.15   \\
J1136409$-$075511  &  LP 673-63   &  1.17   &  1.37   &  0.96   &  1.07   &  0.91   &  1.18   &  0.923   &  0.541   &  0.724   &  0.507   &  2.12   \\
J1141440$-$223215  &  ...         &  0.90   &  2.01   &  1.00   &  1.26   &  1.03   &  1.42   &  0.678:  &  0.245   &  0.467   &  0.188:  &  2.20   \\
J1145354$-$202105  &  LP 793-34   &  1.10   &  1.30   &  0.99   &  1.16   &  0.99   &  1.17   &  0.766   &  0.310   &  0.601   &  0.314   &  2.09   \\
J1147421$+$001506  &  LP 613-50   &  1.32   &  1.22   &  0.96   &  1.11   &  0.94   &  1.20   &  0.772   &  0.313   &  0.621   &  0.295   &  2.06   \\
J1201421$-$273746  &  LP 908-5    &  0.86   &  1.39   &  0.92   &  1.10   &  0.89   &  1.16   &  0.892   &  0.272   &  0.588   &  0.257   &  2.10   \\
J1216101$-$112609  &  LP 734-87   &  1.08   &  1.37   &  0.99   &  1.18   &  0.98   &  1.27   &  0.881   &  0.281   &  0.570   &  0.275   &  2.10   \\
J1223562$-$275746  &  LHS 325a    &  0.91   &  1.42   &  0.95   &  1.10   &  0.94   &  1.18   &  0.869   &  0.254   &  0.582   &  0.233   &  2.12   \\
J1236153$-$310646  &  LP 909-55   &  1.36   &  1.44   &  0.99   &  1.19   &  1.01   &  1.36   &  0.790   &  0.279   &  0.498   &  0.271   &  2.08   \\
J1346460$-$314925  &  LP 911-56   &  0.92   &  1.44   &  0.95   &  1.08   &  0.93   &  1.19   &  0.873   &  0.267   &  0.594   &  0.242   &  2.12   \\
J1357149$-$143852  &  ...         &  1.52   &  1.72   &  1.01   &  1.30   &  1.01   &  1.35   &  0.959   &  0.300   &  0.615   &  0.316   &  2.20   \\
J1406493$-$301828  &  LHS 2859    &  1.17   &  1.29   &  0.96   &  1.04   &  0.92   &  1.12   &  0.862   &  0.348   &  0.662   &  0.322   &  2.05   \\
J1412069$-$041348  &  LP 679-32   &  0.91   &  1.26   &  0.94   &  1.05   &  0.89   &  1.09   &  0.907   &  0.373   &  0.711   &  0.345   &  2.07   \\
J1512333$-$103241  &  ...         &  1.06:  &  1.94   &  0.99   &  1.24   &  1.00   &  1.35   &  0.824:  &  0.246   &  0.650   &  0.340   &  2.29   \\
J1546115$-$251405  &  LP 860-30   &  1.25   &  1.30   &  0.96   &  1.10   &  0.94   &  1.20   &  0.836   &  0.341   &  0.651   &  0.304   &  2.04   \\
J1614252$-$025100  &  LP 624-54   &  1.12   &  1.52   &  0.95   &  1.12   &  0.95   &  1.24   &  0.843   &  0.270   &  0.583   &  0.249   &  2.12   \\
J1645282$-$011228  &  LP 626-2    &  1.22   &  1.39   &  0.97   &  1.09   &  0.96   &  1.20   &  0.852   &  0.296   &  0.577   &  0.271   &  2.06   \\
J2002134$-$542555  &  ...         &  1.25   &  1.48   &  0.96   &  1.13   &  0.98   &  1.27   &  0.783   &  0.257   &  0.562   &  0.250   &  2.10   \\
J2132297$-$051158  &  LP 698-2    &  0.89   &  1.34   &  0.93   &  1.04   &  0.89   &  1.10   &  0.865   &  0.340   &  0.694   &  0.294   &  2.08   \\
J2151270$-$012713  &  LP 638-50   &  1.20   &  1.29   &  0.97   &  1.10   &  0.98   &  1.18   &  0.796   &  0.308   &  0.599   &  0.304   &  2.05   \\
J2205357$-$110428  &  LP 759-25   &  1.10   &  1.41   &  0.97   &  1.12   &  0.98   &  1.23   &  0.819   &  0.276   &  0.562   &  0.280   &  2.08   \\
J2353594$-$083331  &  ...         &  1.33   &  2.07   &  0.99   &  1.27   &  0.96   &  1.45   &  0.876   &  0.348   &  0.679   &  0.356   &  2.29   \\
VB10               &  LHS 474     &  1.09   &  1.87   &  0.96   &  1.14   &  0.92   &  1.32   &  0.939   &  0.318   &  0.633   &  0.307   &  2.32   \\
    \noalign{\smallskip}
    \hline 
   \end{tabular}
  $$
  \begin{list}{}{}
  \item[] 
Columns 1 \& 2 : Full DENIS name with the DENIS-P prefix and NLTT/LHS name.

Columns 3-13: Spectroscopic indices. H$_{\alpha}$ defined in Reid, Hawley \& Gizis (\cite{reid95}); PC3, CrH1, CrH2, FeH1, FeH2
in Mart\'{\i}n et al.(\cite{martin99}); CaH1, CaH2, CaH3, TiO5 in Reid, Hawley \& Gizis  (\cite{reid95}); 
VOa in Kirkpatrick et al. (\cite{kirkpatrick99}). A colon (':') indicates unreliable values due to low signal-to-noise
spectra.

  \end{list}
\end{table*}
\begin{figure*}
\hspace{0.7cm}
\psfig{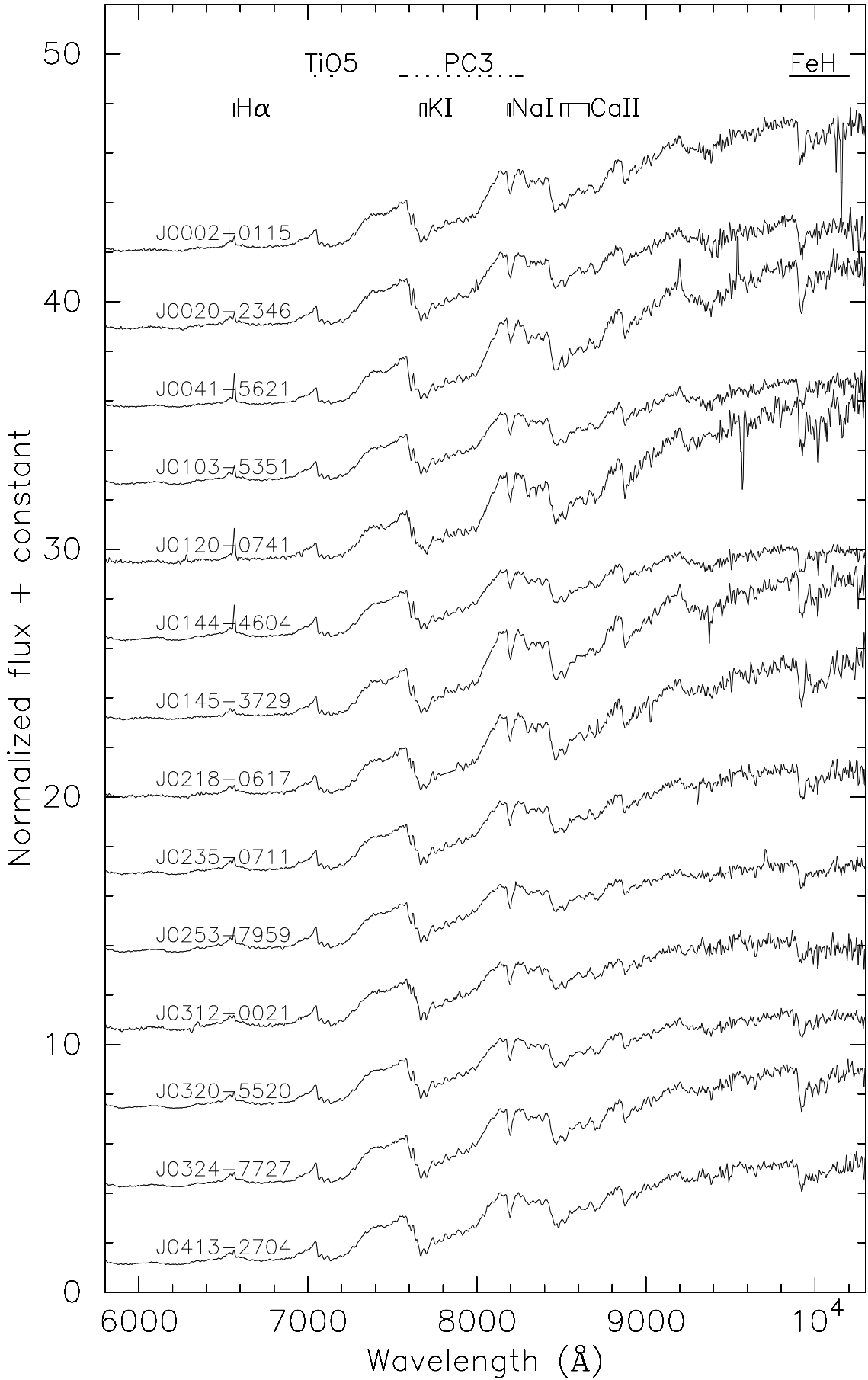}
\caption{Spectra of the 41 late-M dwarfs and VB 10. The positions of 
the H$_{\alpha}$,  Na{\small I}, K{\small I} and Ca{\small II} lines 
are indicated, as well as the spectral intervals used to compute the 
TiO5, and PC3 indices. The FeH absorption features from 985 to 1020~nm are also indicated.}
\label{fig_spectra1}	
\end{figure*}
\begin{figure*}
\setcounter{figure}{0}
\hspace{0.7cm}
\psfig{width=15.0cm,file=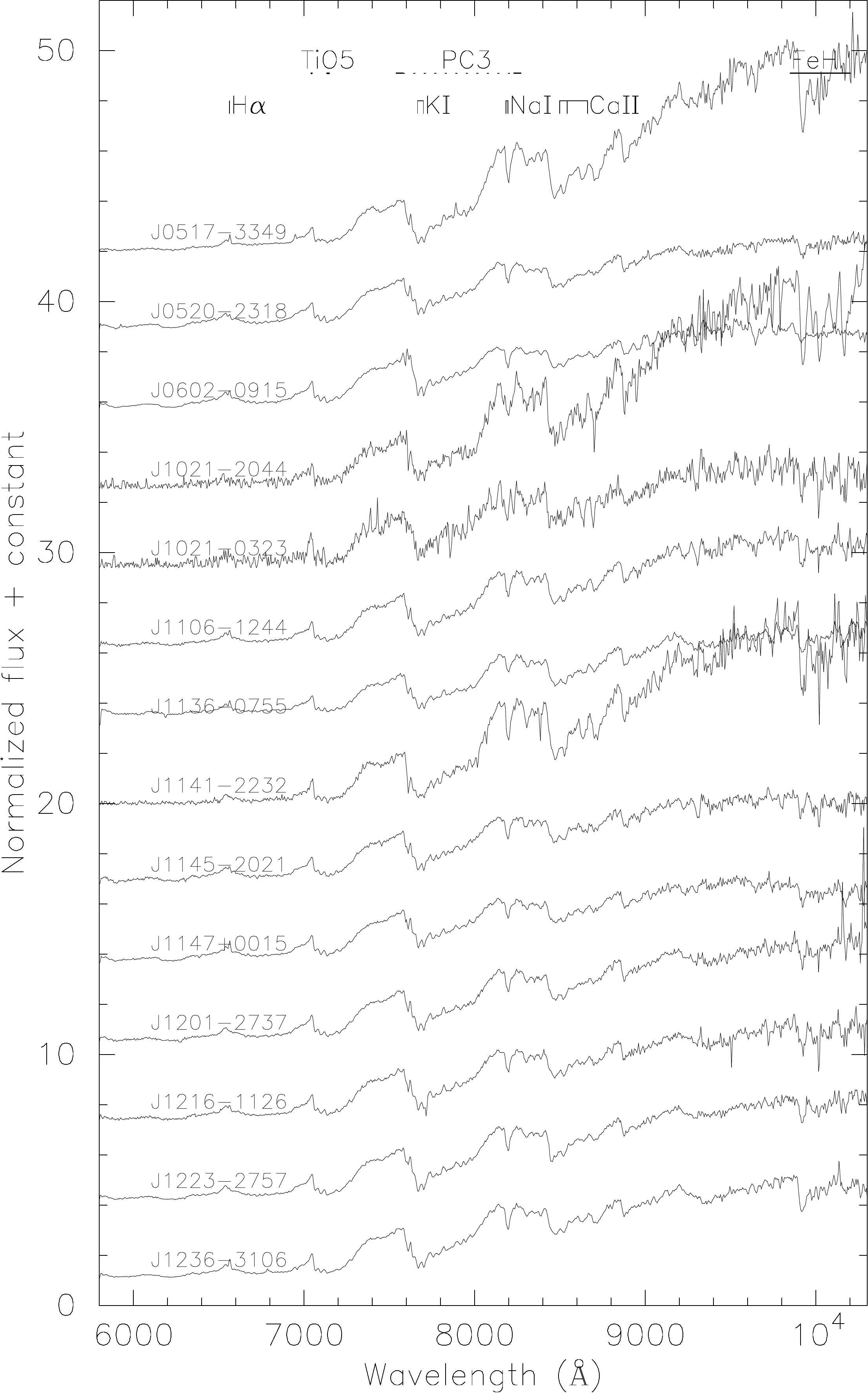,angle=-90}
\caption{continued}
\label{fig_spectra1}	
\end{figure*}
\begin{figure*}
\setcounter{figure}{0}
\hspace{0.7cm}
\psfig{width=15.0cm,file=Figure1c.ps,angle=-90}
\caption{continued}
\label{fig_spectra1}	
\end{figure*}
\begin{figure*}
\psfig{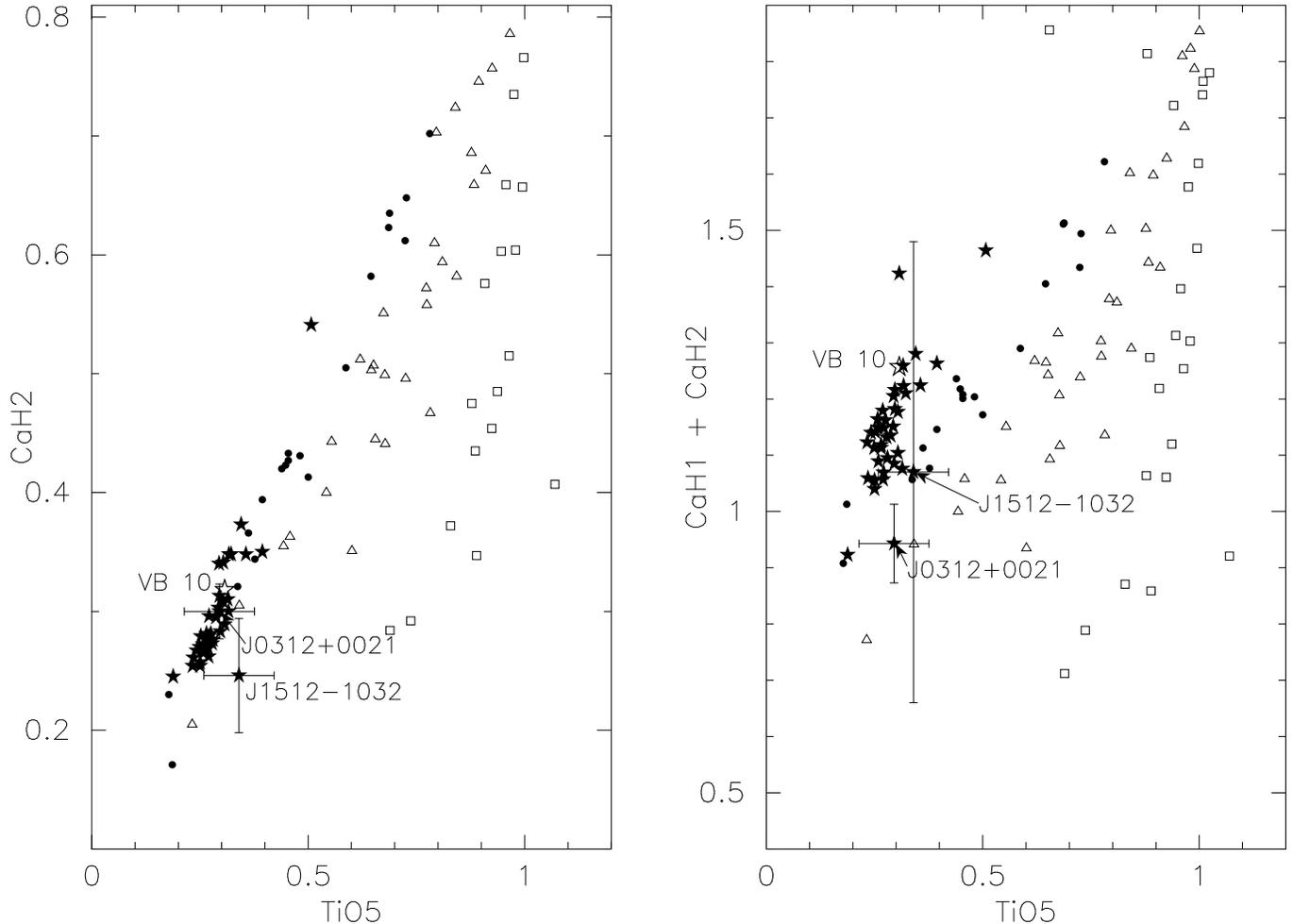}
\caption{CaH vs. TiO5 diagrams. The filled stars represent the measurements of this paper.
Solid circles are M dwarfs, open triangles are sdM subdwarfs and open squares are esdM subdwarfs
from Gizis (\cite{gizis97}). The empty star is VB 10.
Two possible M subdwarfs are indicated: DENIS-P J0312+0021 and J1512$-$1032.
In the right panel DENIS-P J0312$+$0021 appears a probable
M subdwarf due to large error bars, however the CaH1 vs. TiO5 diagram
indicates that DENIS-P J0312$+$0021 is well located on the dwarf sequence and it is ruled out as 
metal poor dwarf. Comparison of spectra between DENIS-P J1512$-$1032 and subdwarf templates also
confirms it to be a late-M dwarf, see Fig.~\ref{fig_subdwarfs}.
A few stars with unreliable spectral indices are not plotted (Table~\ref{table_indices}),
see discussion in Sec.~3.}
\label{fig_indices}	
\end{figure*}

Papers~I and II examined a total of 132 DENIS sources with $2\leq I-J\leq3$,
and classified them into 80 probable dwarfs and 52 probable giants using the MRPM method. 
Crifo et al. (\cite{crifo}) observed 36 of the 80 dwarf candidates and 
confirmed 32 were M dwarfs and 4 were reddened F-K main sequence star contaminants.
Phan-Bao et al. (\cite{phan-bao05}) reported one object LP 714-37 as a binary system of very low mass stars.
Our present sample includes the 43 remaining candidates plus two additional stars from the 52 probable 
giants list (Table 4, Paper II). These two candidates were fainter than the plate limit, 
so we could not measure their proper motion and they were therefore not ruled out by the MRPM. 
On the other hand, as they fell on the dwarf star sequence in the ($I-J$, $J-K$) diagram 
(see Fig.~2, Paper I; Bessell \& Brett \cite{bessell88}) we 
added them to our spectroscopic sample list in order to confirm their nature.
We finally had a total of 45 candidates.

\subsection{Spectroscopic observations and reductions}
During our run, forty two of the stars were observed in July 2005 with the DBS spectrograph on the 2.3m telescope
at Siding Spring Observatory with the 158g/mm grating providing a wavelength coverage of
580 - 1030~nm at 0.5~nm resolution. Three remaining stars 
(LHS 2049, 5165 and LP 859-1) were not been observed due to their having too large an airmass,
however, these stars have previous classifications as M dwarfs in the literature.

The data were reduced using FIGARO. Smooth spectrum stars were observed at a range of airmass to 
remove the telluric lines as described in Bessell (\cite{bessell99}) and the spectrophotometric standard 
EG131 (loc cit) was used to put the data on a relative absolute flux scale. A NeAr arc was 
used for wavelength calibration.  

All spectra were normalized over the 754-758~nm interval that is
the denominator of the PC3 index and a region with a good flat 
pseudo-continuum.
\begin{figure}
\hspace{0.5cm}
\psfig{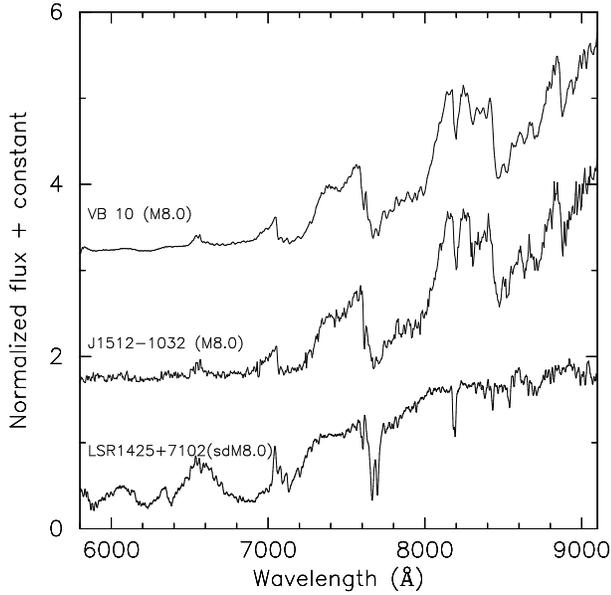}
\caption{Comparison between spectra of DENIS-P J 1512$-$1032 (M8.0) and an M8.0 dwarfs (VB 10, top) and
an M8.0 subdwarf template (LSR 1425+7102, L\'epine, Shara \& Rich \cite{lepine03b}).}
\label{fig_subdwarfs}	
\end{figure}

At the resolution of these spectra, the presence of the Na{\small I} 
and K{\small I} doublets, the presence of FeH and the appearance of
strong CaH cutting into the continuum shortward of 700~nm and the absence of the Ca{\small II} triplet 
immediately distinguish M dwarfs from M giants (see, Bessell \cite{bessell91}).

One of the 42 observed targets, LP 862-111 (or J1641$-$2359), has a much earlier spectrum indicating that it has
been reddened by intervening dust as discussed in Crifo et al. (\cite{crifo}).
Blue spectra would be needed for precise classification at earlier spectral types, but 
from the strength of the Paschen lines the star is probably an F-G dwarf (Figure~\ref{fig_FKdwarfs}).

Figure~\ref{fig_spectra1} shows the 41 spectra of nearby late-M dwarfs and the VB~10 (M8.0) standard obtained
at the MSSSO-2.3m telescope. 
\begin{table*}
   \caption{Estimated absolute magnitude, spectrophotometric distance and spectral type for 
    the 41 nearby M dwarfs and VB 10 (M8.0)}
    \label{table_resultat}
  $$
   \begin{tabular}{lllllllllllllll}
   \hline 
   \hline
   \noalign{\smallskip}
Stars     & $R-I$   &   $M_{I}$ & $M_{J}$ & $M_{K}$ & $d_{I}$  & $d_{J}$ & $d_{K}$  & $d_{sp}$ & d  &  Sp.T & Sp.T  & Sp.T  & Sp.T  & EW\\
          &         &      &    &   &     &    &     &     & (IJ)  & VOa &  TiO5 & PC3   & adopted &H$_{\alpha}$ (\AA) \\
  (1)&(2) &(3) &(4)  &(5) &(6) &(7) &(8)  &(9) &(10) & (11) & (12) & (13) & (14) & (15) \\
   \hline
   \noalign{\smallskip}  
J0002$+$0115          &  2.55  &  13.13  &  10.62  &~\,9.65  &  21.6  &  21.4  &  20.0  &  21.0  &  20.4  &  M6.9  &  M5.3  &  M6.9  &  M6.5  &~~\,9.8   \\
J0020$-$2346          &  2.48  &  12.77  &  10.43  &~\,9.51  &  23.7  &  23.3  &  22.6  &  23.2  &  21.4  &  M6.0  &  M5.0  &  M6.4  &  M6.0  &~~\,8.1   \\
J0041$-$5621          &  2.60  &  13.46  &  10.79  &~\,9.78  &  17.6  &  16.9  &  16.5  &  17.0  &  16.5  &  M7.6  &  M7.7  &  M7.3  &  M7.5  &  ~37.1   \\
J0103$-$5351          &  2.41  &  12.49  &  10.26  &~\,9.38  &  25.8  &  25.2  &  21.9  &  24.3  &  23.1  &  M6.0  &  M5.0  &  M6.0  &  M5.5  &~~\,9.8   \\
J0120$-$0741          &  2.57  &  13.84  &  11.00  &~\,9.93  &  23.7  &  24.5  &  23.1  &  23.8  &  26.0  &  M8.0  &  M8.0  &  M7.8  &  M8.0  &  ~51.3   \\
J0144$-$4604          &  2.38  &  12.23  &  10.11  &~\,9.25  &  23.7  &  22.9  &  23.4  &  23.3  &  20.8  &  M5.6  &  M5.3  &  M5.6  &  M5.5  &  ~25.4   \\
J0145$-$3729          &  2.62  &  13.67  &  10.91  &~\,9.86  &  18.9  &  20.7  &  21.7  &  20.5  &  22.3  &  M7.2  &  M7.6  &  M7.6  &  M7.5  &~~\,4.8   \\
J0218$-$0617$^{\rm c}$&  2.57  &  13.27  &  10.70  &~\,9.71  &  28.6  &  27.7  &  24.9  &  27.1  &  26.4  &  M7.1  &  M7.6  &  M7.1  &  M7.0  &~~\,6.0   \\
J0235$-$0711          &  2.41  &  12.61  &  10.34  &~\,9.43  &  26.3  &  25.8  &  24.5  &  25.5  &  23.6  &  M5.9  &  M5.1  &  M6.2  &  M5.5  &~~\,7.2   \\
J0253$-$7959          &  2.36  &  12.23  &  10.11  &~\,9.25  &  17.7  &  17.4  &  16.5  &  17.2  &  16.4  &  M5.3  &  M5.3  &  M5.6  &  M5.5  &  ~12.4   \\
J0312$+$0021          &  2.34  &  12.18  &  10.08  &~\,9.23  &  26.5  &  25.6  &  24.8  &  25.7  &  23.4  &  M5.8  &  M5.0  &  M5.6  &  M5.5  &~~\,3.9   \\
J0320$-$5520          &  2.38  &  12.40  &  10.21  &~\,9.34  &  24.0  &  23.5  &  21.9  &  23.1  &  21.7  &  M5.7  &  M5.0  &  M5.9  &  M5.5  &~~\,7.0   \\
J0324$-$7727          &  2.49  &  12.97  &  10.54  &~\,9.59  &  19.0  &  20.6  &  19.7  &  19.7  &  21.5  &  M5.6  &  M5.4  &  M6.7  &  M6.0  &~~\,4.7   \\
J0413$-$2704$^{\rm d}$&  2.42  &  12.44  &  10.24  &~\,9.36  &  25.2  &  24.6  &  23.4  &  24.4  &  22.4  &  M5.9  &  M5.5  &  M5.9  &  M6.0  &~~\,5.2   \\
J0517$-$3349          &  2.62  &  14.70  &  11.52  &  10.34  &  11.1  &  12.7  &  12.6  &  12.1  &  16.4  &  M7.1  &  M8.5  &  M8.9  &  M8.0  &~~\,8.1   \\
J0520$-$2318          &  2.26  &  11.91  &~\,9.92  &~\,9.08  &  26.4  &  23.3  &  20.4  &  23.4  &  18.2  &  M5.4  &  M4.8  &  M5.2  &  M5.0  &~~\,2.1   \\
J0602$-$0915$^{\rm c}$&  2.26  &  11.39  &~\,9.57  &~\,8.76  &  20.8  &  18.1  &  17.8  &  18.9  &  13.7  &  M5.5  &  M5.2  &  M4.5  &  M5.0  &~~\,4.4   \\
J1021$-$2044          &  2.60  &  14.64  &  11.48  &  10.31  &  19.5  &  21.2  &  21.8  &  20.8  &  25.6  &  M7.2  &  M7.7: &  M8.9  &  M8.0  &$<$2.9   \\
J1021$-$0323$^{\rm d}$&  2.39  &  12.40  &  10.21  &~\,9.34  &  26.9  &  25.4  &  25.4  &  25.9  &  22.2  &  M4.3: &  M7.3: &  M5.9  &  M6.0  &$<$3.0   \\
J1106$-$1244          &  2.46  &  12.53  &  10.29  &~\,9.40  &  21.4  &  19.8  &  20.5  &  20.5  &  16.9  &  M6.3  &  M5.4  &  M6.0  &  M6.0  &~~\,6.4   \\
J1136$-$0755          &  1.99  &  12.18  &  10.08  &~\,9.23  &  26.6  &  25.4  &  24.9  &  25.7  &  22.6  &  M6.0  &  M2.7: &  M5.6  &  M6.0  &~~\,5.3   \\
J1141$-$2232          &  2.69  &  14.50  &  11.38  &  10.22  &  15.3  &  18.3  &  17.9  &  17.2  &  23.6  &  M6.9  &  M7.3: &  M8.7  &  M8.0  &$<$0.8   \\
J1145$-$2021          &  2.29  &  11.87  &~\,9.89  &~\,9.06  &  24.8  &  22.9  &  22.4  &  23.4  &  19.5  &  M5.7  &  M4.8  &  M5.1  &  M5.0  &~~\,3.3   \\
J1147$+$0015          &  2.26  &  11.48  &~\,9.64  &~\,8.83  &  21.9  &  19.9  &  18.6  &  20.1  &  16.5  &  M5.4  &  M5.0  &  M4.6  &  M5.0  &~~\,6.8   \\
J1201$-$2737$^{\rm c}$&  2.41  &  12.27  &  10.14  &~\,9.27  &  25.5  &  24.6  &  24.1  &  24.7  &  22.2  &  M5.8  &  M5.4  &  M5.7  &  M5.5  &$<$0.6   \\
J1216$-$1126          &  2.40  &  12.18  &  10.08  &~\,9.23  &  32.2  &  29.3  &  26.2  &  29.2  &  24.3  &  M5.8  &  M5.2  &  M5.6  &  M5.5  &~~\,3.8   \\
J1223$-$2757          &  2.46  &  12.40  &  10.21  &~\,9.34  &  22.8  &  21.6  &  21.6  &  22.0  &  19.0  &  M6.0  &  M5.7  &  M5.9  &  M6.0  &$<$0.9   \\
J1236$-$3106          &  2.42  &  12.49  &  10.26  &~\,9.38  &  19.8  &  19.7  &  18.6  &  19.4  &  18.6  &  M5.6  &  M5.3  &  M6.0  &  M5.5  &~~\,9.3   \\
J1346$-$3149          &  2.45  &  12.49  &  10.26  &~\,9.38  &  14.4  &  14.2  &  13.1  &  13.9  &  13.3  &  M6.0  &  M5.6  &  M6.0  &  M6.0  &$<$0.5   \\
J1357$-$1438          &  2.55  &  13.56  &  10.85  &~\,9.82  &  25.0  &  25.2  &  23.9  &  24.7  &  25.6  &  M6.9  &  M8.0  &  M7.5  &  M7.5  &~~\,9.2   \\
J1406$-$3018$^{\rm c}$&  2.25  &  11.82  &~\,9.86  &~\,9.03  &  20.9  &  19.3  &  17.8  &  19.3  &  16.5  &  M5.3  &  M4.7  &  M5.1  &  M5.0  &~~\,4.1   \\
J1412$-$0413$^{\rm c}$&  2.20  &  11.68  &~\,9.77  &~\,8.95  &  24.9  &  23.5  &  21.8  &  23.4  &  21.2  &  M5.5  &  M4.5  &  M4.9  &  M5.0  &$<$0.6   \\
J1512$-$1032          &  2.63  &  14.29  &  11.25  &  10.12  &  22.0  &  23.4  &  22.9  &  22.8  &  26.6  &  M7.8  &  M8.1  &  M8.4  &  M8.0  &$<$5.0   \\
J1546$-$2514          &  2.25  &  11.87  &~\,9.89  &~\,9.06  &  27.8  &  26.5  &  25.8  &  26.7  &  23.8  &  M5.2  &  M4.9  &  M5.1  &  M5.0  &~~\,6.1   \\
J1614$-$0251          &  2.50  &  12.81  &  10.45  &~\,9.52  &  14.6  &  15.1  &  14.3  &  14.6  &  14.8  &  M6.0  &  M5.5  &  M6.4  &  M6.0  &~~\,4.2   \\
J1645$-$0112          &  2.38  &  12.27  &  10.14  &~\,9.27  &  25.2  &  25.1  &  24.3  &  24.9  &  24.2  &  M5.4  &  M5.3  &  M5.7  &  M5.5  &~~\,5.3   \\
J2002$-$5425          &  2.46  &  12.65  &  10.36  &~\,9.45  &  17.7  &  18.5  &  16.3  &  17.5  &  18.6  &  M5.8  &  M5.5  &  M6.2  &  M6.0  &~~\,7.6   \\
J2132$-$0511$^{\rm c}$&  2.29  &  12.05  &  10.00  &~\,9.16  &  19.7  &  19.0  &  16.7  &  18.5  &  17.5  &  M5.6  &  M5.0  &  M5.4  &  M5.5  &$<$1.1   \\
J2151$-$0127$^{\rm c}$&  2.26  &  11.82  &~\,9.86  &~\,9.03  &  19.0  &  18.5  &  18.6  &  18.7  &  17.8  &  M5.3  &  M4.9  &  M5.1  &  M5.0  &~~\,5.1   \\
J2205$-$1104$^{\rm c}$&  2.38  &  12.36  &  10.19  &~\,9.32  &  18.3  &  18.6  &  17.7  &  18.2  &  18.5  &  M5.6  &  M5.2  &  M5.8  &  M5.5  &~~\,3.6   \\
J2353$-$0833          &  2.64  &  14.67  &  11.50  &  10.32  &  17.8  &  19.6  &  20.0  &  19.1  &  24.0  &  M7.8  &  M8.2  &  M8.9  &  M8.5  &  ~10.1   \\
VB 10                 &  2.62  &14.07$^{\rm b}$&11.12$^{\rm b}$&10.02$^{\rm b}$&~\,5.7  &~\,5.8  &~\,5.7  &~\,5.7$^{\rm a}$&~\,6.3  &  M8.1  &  M8.0  &  M8.1  &  M8.0  &~~\,3.8   \\
    \noalign{\smallskip}
    \hline 
   \end{tabular}
  $$
  \begin{list}{}{}
  \item[$^{\rm a}$]: d$_{\pi}$ = 5.87~pc, derived from $\pi$ = 170.3~mas, a companion of HIP 94761 
  \item[$^{\rm b}$]: optical and infrared photometry from Bessell (\cite{bessell91})
  \item[$^{\rm c}$]: also listed by Reid et al. (\cite{reid03})
  \item[$^{\rm d}$]: see Cruz et al. (\cite{cruz03})
  
Column 1: Abbreviated DENIS name.

Column 2: $R-I$ color computed from spectra.

Column 3, 4 \& 5:  Absolute magnitudes for the $I$, $J$, $K$ bands 
based on the PC3-absolute magnitudes relation.

Columns 6, 7 \& 8: Distance (pc) estimated from the DENIS photometry and
the $M_{\rm I}$, $M_{\rm J}$, $M_{\rm K}$ derived from the PC3 index.
Columns 9 \& 10: Adopted distance, and distance previously derived in paper~II
from the $I-J$ colour.

Column 11, 12, 13 \& 14: Spectral types derived from the VOa, TiO5, and PC3 index
using the formula given in Cruz \& Reid (\cite{cruz}) and Mart\'{\i}n et al. (\cite{martin99}),
and the adopted spectral rounded to the nearest half subtype. A colon (':') indicates unreliable values
which will not be taken into account for the final spectra types.

Column 15: H$_{\alpha}$ equivalent widths (\AA).

  \end{list}
\end{table*}
\begin{figure}
\hspace{0.1cm}
\psfig{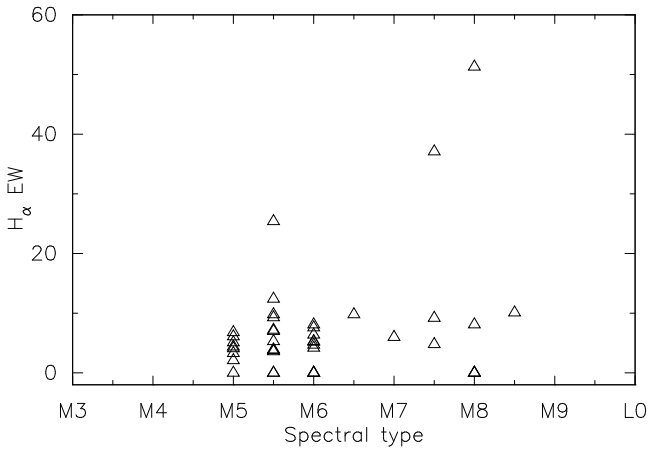}
\caption{H$_{\alpha}$ equivalent width vs. spectral type diagram for all stars with spectral
types computed in this paper.}
\label{fig_EW}	
\end{figure}

\section{Spectroscopic indices, spectral type classifications and distances}

In the past decade, many important spectral indices have been defined for 
M and L dwarfs (e.g., Reid, Hawley \& Gizis \cite{reid95}; Mart\'{\i}n et al. \cite{martin99}; Kirkpatrick
et al. \cite{kirkpatrick99}; L\'epine, Rich \& Shara \cite{lepine03a}). 
These indices allow us to quantify spectral types (e.g., PC, TiO, and VO)
and metallicity classes (e.g., CaH) of red dwarfs.

For the work here, we have used some available indices to estimate spectral types and distinguish between M 
dwarfs and (e)subdwarfs in our sample.
Table~\ref{table_indices} lists our spectral indices measurement. For classifying
spectral types of M dwarfs, we use the PC3 index defined by Mart\'{\i}n et al. (\cite{martin99}),
TiO5 in Reid, Hawley \& Gizis (\cite{reid95}) and VOa in Kirkpatrick et al. (\cite{kirkpatrick99}).
The final spectral type is an average value of three spectral types computed from these three indices,
except some cases as discussed below.

Fig.~\ref{fig_indices} shows the comparison between the CaH and TiO5 indices measured from our 
observations of DENIS nearby candidates and also plots dwarfs and metal-poor dwarfs from Gizis (\cite{gizis97}).
Two stars: J0312$+0021$ and J1512$-$1032, are very faint with low signal-to-noise spectra and could
be mild subdwarfs. However, the combination of two diagrams: TiO5 vs. CaH2 and TiO5 vs. (CaH1+CaH2), 
does not support J0312$+0021$ being a subdwarf, and comparison of the spectra of J1512$-$1032, the standard VB 10 
and an M8.0 subdwarf template (LSR 1425+7102) from S\'ebastien L\'epine clearly shows that J1512$-$1032 
is a late-M dwarf rather than a subdwarf (Fig.~\ref{fig_subdwarfs}).

In fact, comparison of all our spectra with M subdwarf references (e.g., Bessell \cite{bessell82}, Gizis \cite{gizis97}) 
does not reveal any subdwarfs in our sample. This is not unexpected, as extreme subdwarfs are intrinsically very 
faint and rare and our volume size is limited and too small to have much probability of containing extreme halo members. 

Some extreme M subdwarf spectra are given in Bessell (\cite{bessell82}),   
amongst which are the faint pair LHS 2099 (R$\sim$16) and LHS 2100 (R$\sim$19). The spectrum of LHS 2099 
is identical to that of LHS 541 (R$\sim$15), the M subdwarf companion of the well known [Fe/H]=-1.7 subdwarf HD219617 (R$\sim$8).
LHS 2100 is the best example of a very cool [Fe/H]=-1.7 M subdwarf.

Table~\ref{table_resultat} lists our spectral type estimates by using the formulae given in 
Mart\'{\i}n et al. (\cite{martin99}) for the PC3 index and Cruz \& Reid (\cite{cruz}) for the TiO5 and VOa indices.
To avoid saturation in TiO5 around about spectral type M7, we firstly computed spectral types
using the PC3 index. We then used Cruz \& Reid's formula: $S_{p}$ = -10.775(TiO5) + 8.200 for spectral types later than M7
and $S_{p}$ = 5.673(TiO5) + 6.221 for earlier ones.
We adopted the mean spectral types computed from these three indices with an uncertainty of $\pm$0.5 subclass.
One should note for J1136$-$0755 that our spectral type estimate from TiO5 is M2.7, however
comparison of its spectrum with early-M dwarf templates in Bessell (\cite {bessell91})  
indicates clearly that this is a mid-M dwarf rather than an early-M dwarf. 
This is consistent with spectral types estimated from PC3 and VOa and we finally adopted a spectral type of M6.0
for J1136$-$0755 (or LP 673-63).
In this paper, we found three M7.5 dwarfs: J0041$-$5621, J0145$-$3729, J1357$-$1438; five M8.0:
J0120$-$0741, J0517$-$3349, J1021$-$2044, J1141$-$2232, J1512$-$1032; one M8.5: J2353$-$0833, the latest in our 
sample is at 19~pc. All of them have proper motions measured in Paper~II, except J1021$-$2044 and J1512$-$1032
that were below the plate limit plate and are located in the 25~pc volume.

To estimate the distances, we used the magnitude versus PC3 index relations given in Crifo et al. (\cite{crifo})
with a $\sim$12\% error in the distance to single stars.
Table~\ref{table_resultat} lists 
the absolute magnitudes in the three DENIS bands computed from the
PC3 index, as well as the estimated spectrophotometric distances for 
each of those bands and their average. The values for the three colours $I$, 
$J$, $K$ are very similar, indicating correlated uncertainties for the three
estimators. For unresolved binaries in our sample, their distances are 
underestimated by up to $\sqrt{2}$.
We also list in the table our distances estimated from the $I-J$ color to I-band absolute magnitude
relation given in Paper II. The distances computed from 4 estimators are reasonably well matched.
In this table, there are two new M8.0 members of the 25~pc volume: J1021$-$2044 and J1512$-$1032.
\begin{figure}
\hspace{0.2cm}
\psfig{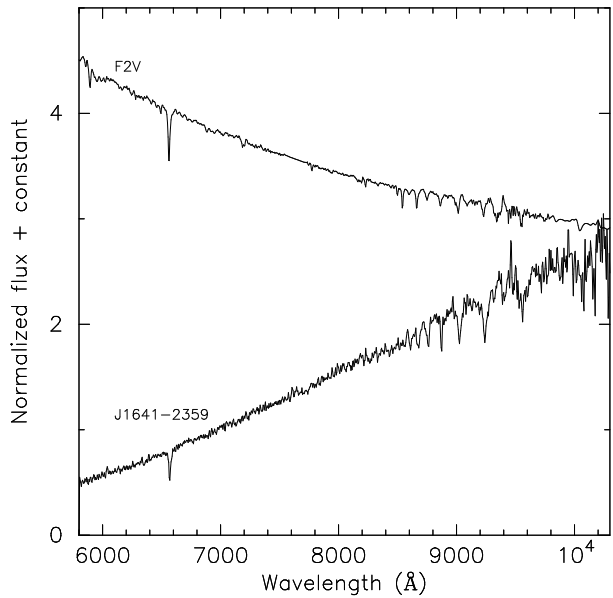}
\caption{Comparison between spectra of an F2V dwarf (top) from Pickles (\cite{pickles})
and DENIS-P J 1614$-$2359 reddened by intervening dust (bottom).}
\label{fig_FKdwarfs}	
\end{figure}

\section{Discussion}
Seventy eight percent of the stars in our sample exhibited H$_{\alpha}$ emission.  
We directly measured H$_{\alpha}$ equivalent widths using the IRAF task SPLOT. 
Table~\ref{table_resultat} (column 15) lists our measurements for 41 stars.
We also measured an upper limit for the remaining stars which have a weak H$_{\alpha}$ emission or 
low signal-to-noise spectra.
Note that there are two late-M dwarfs exhibiting very strong H$_{\alpha}$ emission: J0041$-$5621 (M7.5, EW = 37.1) and
J0120$-$0741 (M8.0, EW = 51.3).
 
It is interesting to note that a flaring state was seen in the spectrum of LP~890-2 (or J0413$-$2704, M6.0)
obtained by Cruz \& Reid (\cite{cruz}) in comparison with ours.
This M6.0 dwarf showed strong  H$_{\alpha}$ emission (EW = 16.0~\AA) in
Cruz and Reid's observation (their Fig.~7) but is much weaker in ours (EW = 5.0~\AA).
Fig.~\ref{fig_EW} plots H$_{\alpha}$ equivalent widths versus spectral types. 
One M8.0 dwarf (J1141-2232) has no emission or too weak H$_{\alpha}$ emission, we set an upper limit of 0.8~\AA~
for this late-M star.
 
We note that one of our targets DENIS-P J1614$-$2359 (LP 826-111) is a reddened F-G main sequence star. 
Fig.~\ref{fig_FKdwarfs} presents comparison between spectra of DENIS-P J1614$-$2359
and an F2V main sequence star from Pickles (\cite{pickles}).
This object is probably in 
the background of the Rho Ophiuchus molecular complex and reddened by intervening dust as discussed in
Crifo et al. (\cite{crifo}). 

\section{Summary}
We have presented updated spectroscopic follow-up observations for the DENIS nearby star candidates detected
in Paper~I and II. Forty one M dwarfs are spectroscopically confirmed in this paper, 
Crifo et al. (\cite{crifo}) also presented 32 M dwarfs. 
All these stars were pinpointed using the MRPM method that is a robust tool for searching for
new nearby ultracool and brown dwarfs in both low and high proper motion surveys in the optical-infrared.
\begin{acknowledgements}
This research is carried out based on the DENIS photometry kindly provided by the DENIS consortium.
P-B.N. is grateful to Guy Simon for help during the work.
We thank S\'ebastien L\'epine for kindly providing his published spectra.
We also thank the referee for many useful comments that clarified and much improved our paper.
This research has made use of the SIMBAD and VIZIER databases, operated at CDS, Strasbourg, France.
\end{acknowledgements}


\begin{thebibliography}{}

 
\bibitem[1982]{bessell82} 
  Bessell, M.S. 1982, PASA, 4, 417
  
\bibitem[1988]{bessell88} 
  Bessell, M.S., \& Brett, J.M. 1988, PASP, 100, 1134

\bibitem[1991]{bessell91} 
  Bessell, M.S. 1991, AJ, 101, 662

\bibitem[1999]{bessell99} 
  Bessell, M.S. 1999, PASP, 111, 1433

\bibitem[2005]{crifo}
  Crifo, F., Phan-Bao, N., Delfosse, X. et al. 2005, astro-ph/0506365

\bibitem[2002]{cruz}
  Cruz, K.L., \& Reid, I.N. 2002, AJ, 123, 2828
  
\bibitem[2003]{cruz03}
  Cruz, K.L., Reid, I.N., Liebert, J., et al. 2003, AJ, 126, 2421

\bibitem[1997]{epchtein97} Epchtein, N. 1997, in 
the 2nd DENIS Euroconference,
The impact of large scale near-infrared surveys, 
ed. F. Garzon et al. (Kluwer Dordrecht), 15

\bibitem[1997]{gizis97}
  Gizis, J.E. 1997, AJ, 113, 806
  
\bibitem[1999]{kirkpatrick99}
  Kirkpatrick, J.D., Reid, I.N., Liebert, J., et al. 1999, ApJ, 519, 802
  
\bibitem[2003]{lepine03a}
  L\'epine, S., Rich, R.M., \& Shara, M.M.  2003, AJ, 125, 1598
  
\bibitem[2003]{lepine03b}
  L\'epine, S., Shara, M.M., \& Rich, R.M. 2003, ApJ, 585, L69

\bibitem[1979]{luytena} 
  Luyten, W.J. 1979, Catalogue of stars with proper motions exceeding 
   0\arcsec.5 annually (LHS) (Minneapolis, University of Minnesota)

\bibitem[1980]{luytenb} 
  Luyten, W.J. 1980, New Luyten catalog of stars with proper motions 
  larger than Two Tenths of an arcsecond (NLTT) (Minneapolis, University of Minnesota)
  
\bibitem[1999]{martin99} 
  Mart\'{\i}n, E. L., Delfosse, X., et al. 1999, AJ, 118, 2466

\bibitem[2001]{phan-bao01}
  Phan-Bao, N., Guibert, J., Crifo, F., Delfosse, X., Forveille, T., et al. 2001, A\&A, 380, 590
  
\bibitem[2003]{phan-bao03}
  Phan-Bao, N., Crifo, F., Delfosse, X., Forveille, T., Guibert, J., et al. 2003, A\&A, 401, 959

\bibitem[2005]{phan-bao05}
  Phan-Bao, N., Mart\'{\i}n, E.L., Reyl\'e, C., et al. 2005, A\&A, 439, L19
  
\bibitem[1998]{pickles}
  Pickles, A.J. 1998, PASP, 110, 863

\bibitem[1995]{reid95} 
  Reid, I.N., Hawley, S.L., \& Gizis, J.E. 1995, AJ, 110, 1838

\bibitem[2003]{reid03} 
  Reid, I.N., Cruz, K.L., Allen, P., et al. 2003, AJ, 126, 3007
  
 
\end{thebibliography}
\end{document}